# Enhancing Wildlife Density Estimation: A New Two-Parameter Detection Function for Line Transect Sampling

**Midhat M. Edous**
College of medicine, Alfaisal University, Riyadh 11533, Saudi Arabia
medoos@alfaisal.edu

**Omar M. Eidous**
Department of Statistics, Yarmouk University, Irbid 21162, Jordan
omarm@yu.edu.jo

## Abstract

Accurate estimation of wildlife density is vital for effective ecological monitoring, conservation, and management. Line transect sampling, a central technique in distance sampling, relies on selecting an appropriate detection function to model the probability of detecting individuals as a function of their distance from the transect line. In this study, we propose a novel two-parameter detection function that extends the flexibility of traditional models such as the half-normal and exponential, while retaining interpretability and computational tractability. Notably, one of the parameters is assumed to take a known integer value, allowing us to explore a range of detection curve shapes by varying this parameter across different settings in our computational analysis. This structure enables the model to capture a broader spectrum of detection patterns, especially in cases where classical models fall short. The proposed method is evaluated through extensive simulation studies and applied to real ecological survey data. The results show that the new model consistently yields improved fit and more accurate estimates of animal density, offering ecologists a practical and robust alternative for use in diverse field conditions

# 1. Introduction

Line transect sampling is a widely used, cost-efficient method for estimating the population density of objects, such as animals or plants, within a specified area. Favored by ecologists for its practicality and straightforward implementation, this technique involves traversing a defined transect line while recording both the number of detections and their perpendicular distances from the line (i.e., the shortest distance between each detected object and the observer's path). A key feature of this method is the use of a detection function, which models the likelihood of detecting an object as a function of its distance from the transect. When sampling is conducted randomly, the probability density function (PDF) of observed distances mirrors the shape of the detection function, adjusted to integrate to one. Seminal contributions to the development of this distance sampling framework can be found in the foundational works of [1, 2].

A central assumption in line transect sampling is that the detection function, denoted $g(x)$, decreases monotonically with increasing distance $X$ from the transect line. Specifically, it is assumed that $g(0) = 1$, indicating perfect detectability at the transect line. Moreover, the condition $g'(0) = 0$ is typically imposed to ensure a "shoulder" near the transect line, representing a region of high detectability close to the observer. As a result, the associated PDF of the observed perpendicular distances is also monotonically decreasing.

Burnham and Anderson [3] formulated a foundational expression for estimating the density $D$ of objects within a survey area. This expression is given by,

$$D = \frac{E(n)f(0)}{2L},$$

where $E(n)$ is the expected number of detections, $L$ is the total length of the transect lines, and $f(0)$ is the value of the probability density function of perpendicular distances evaluated at zero. Under standard assumptions, this leads to the commonly used estimator,

$$\widehat{D} = \frac{n\hat{f}(0)}{2L}.$$

Here, $n$ is the observed number of detections, and $\hat{f}(0)$ is an estimator of $f(0)$ derived from the sample of observed perpendicular distances. Therefore, accurately modeling the detection function and obtaining a reliable estimate of $f(0)$ are crucial steps in line transect analysis. Let $f(x)$ represent the unknown PDF of the perpendicular distances $X$, which are typically modeled as independent and identically distributed random variables $X_1, X_2, \ldots, X_n$ (see [2]). A parametric approach assumes that $f(x)$ belongs to a known family of distributions characterized by a parameter $\theta$, which may be scalar or vector-valued. This parameter is estimated using the observed data. Once $\theta$ is estimated, it enables the computation of $f(0)$, which in allows for the estimation of population density $D$.

Several methods are available for estimating the parameter $\theta$, which in turn allows for the estimation of $f(0)$ and population density $D$. Two primary approaches are commonly used: parametric and nonparametric methods. In the parametric approach, it is assumed that the functional form of $f(x)$ is known up to one or more parameters. Popular models include the exponential and half-normal distributions. The exponential model, introduced by Gates et al. [4], has a detection function $g(x;\theta) = e^{-x/\theta}$ and leads to a simple maximum likelihood estimator (MLE) for $f(0)$: $\hat{f}_{EX}(0) = 1/\bar{X}$, though it does not satisfy the shoulder condition. In contrast, the half-normal model, introduced by Hemingway [5], satisfies the shoulder condition and generally provides better performance. Its detection function is $g(x;\sigma) = e^{-x^2/2\sigma^2}$, and the MLE for $f(0)$ is $\hat{f}_{HN}(0) = \sqrt{2n}/\sqrt{\pi \sum_{i=1}^{n} X_i^2}$, see [6, 7].

Numerous alternative parametric models have been proposed (see [8–19, 44]). In addition, nonparametric methods, such as histogram and kernel estimators, are widely used for modeling detection functions without assuming a specific functional form (see [20–32]). Semi-parametric approaches, which blend parametric structure with nonparametric flexibility, have also been explored (see [33–41]).

In this study, we introduce a new two-parameter model for line transect data, which is reduced to a single-parameter form by fixing one parameter at a specified value. Based on this model, several estimators for $f(0)$ are developed, and their performance, particularly in small-sample settings, is assessed and compared to classical estimators using simulation.

## 2. Proposed Detection Function

We introduce a new detection function defined as,
$$g(x; m, \beta) = (1 + \beta x)^m e^{-m\beta x}, \quad x \geq 0, \quad \beta > 0, \quad m = 1,2,3,\ldots$$
This function incorporates two parameters, $m$ and $\beta$, which provide additional flexibility in shaping the decay behavior of detectability over distance. The parameter $\beta$ controls the overall decay rate, while $m$ adjusts the curvature, enabling the function to model a broader range of detection scenarios. To ensure the function meets the shoulder condition, which requires the slope of the detection function to be zero at the origin, we compute the first derivative of $g(x; m, \beta)$ with respect to $x$,
$$g'(x; m, \beta) = -m\beta e^{-m\beta x}(1 + \beta x)^m + m\beta e^{-m\beta x}(1 + \beta x)^{m-1}.$$
Evaluating this derivative at $x = 0$ gives,

$$g'(0; m, \beta) = -m\beta + m\beta = 0.$$

Thus, the proposed function satisfies the shoulder condition. To analyze the monotonic behavior of $g(x; m\,\beta)$, we can factor the first derivative as,

$$g'(x; m, \beta) = m\beta e^{-m\beta x}(1 + \beta x)^{m-1}(-\beta x).$$

Since $m > 0, \beta > 0, e^{-m\beta x} > 0$ and $(1 + \beta x)^{m-1} > 0$, the sign of the derivative is determined by $-\beta x$, which is non-positive for $x \geq 0$. Therefore, $g(x; m\,\beta)$ is a monotonically decreasing function in $x$.

At distance $x = 0$, the detection probability $g(0; m, \beta) = 1$, indicating perfect detection on the transect line. Figure (1) compares three detection functions: the exponential (EX) ($\theta = 1$), half-normal ($\sigma^2 = 1$) and the proposed model ($m = 5$, $\beta = 0.4$). Figures 2 and 3 show the behavior of the proposed detection function for $m = 1$ and $m = 5$, respectively, across different values of $\beta$ ($\beta = 0.25,\ 0.35,\ 0.55,\ 0.85$).

In contrast to the exponential function, which drops sharply as distance increases from $x = 0$, the proposed detection function provides a more gradual and tunable decline. Compared to the half-normal function, which has a fixed decay, the proposed model offers enhanced flexibility by adjusting the shape of the curve through parameter $\beta$. As demonstrated in Figures 2 and 3, adjusting $\beta$ alters the detection range, enabling adaptation to varying field conditions.

Figures (2) and (3) demonstrate the effect of parameter mmm on the shape of the proposed detection function. With $m = 1$ (Figure 2), the curves decline gradually, resulting in broader detection ranges, suitable for detecting objects farther from the transect line, especially at lower $\beta$ values. In contrast, $m = 5$ (Figure 3) produces steeper curves, concentrating detection near the line. Higher mmm values lead to more localized and sensitive detection, making them ideal for precise or densely populated environments.

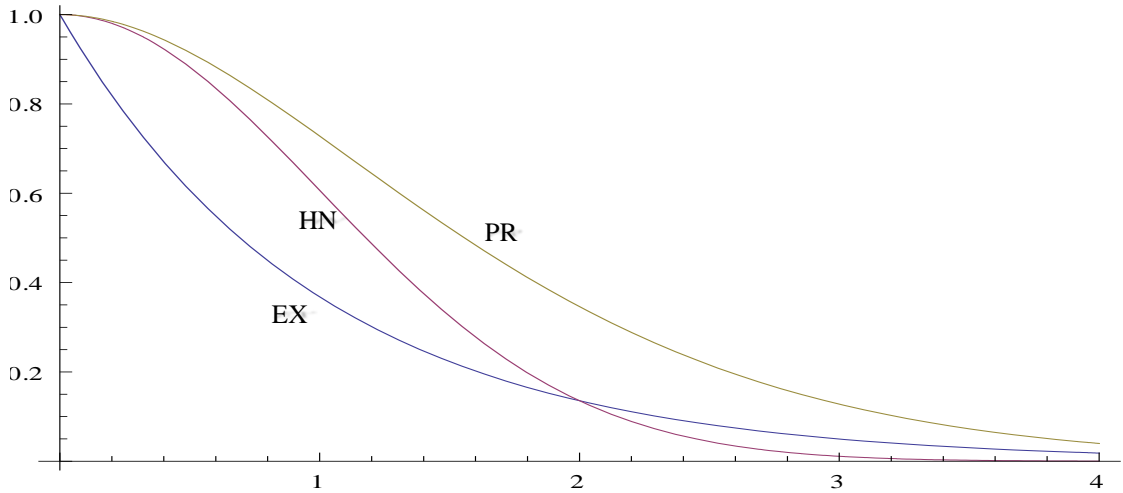

**Figure (1).** Detection functions for exponential model (EX, $\theta = 1$), half-normal model (HN, $\sigma^2 = 1$) and the proposed model (PR, $m = 5$, $\beta = 0.4$).

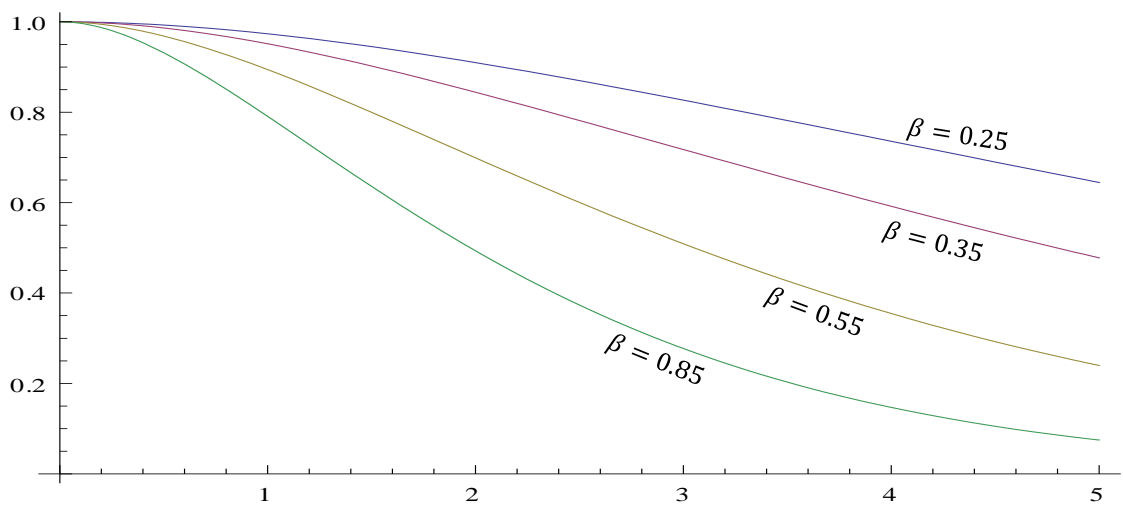

**Figure (2).** Proposed detection function curves with $m = 1$ and varying $\beta$ values: 0.25, 0.35, 0.55 and 0.85.

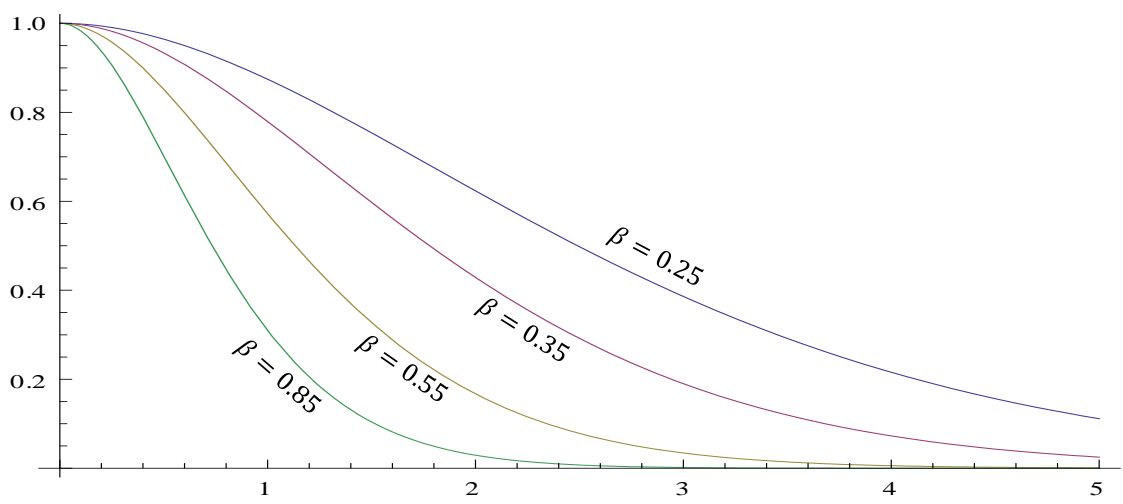

**Figure (3).** Proposed detection function curves with $m = 5$ and varying $\beta$ values: 0.25, 0.35, 0.55 and 0.85.

## 3. PDF Corresponding to the Proposed Detection Function

Given the proposed detection function,
$$g(x; m, \beta) = (1 + \beta x)^m e^{-m\beta x},$$
the corresponding probability density function (PDF) is defined as,
$$f(x; m\, \beta) = C \cdot g(x; m, \beta), \qquad x \geq 0, \qquad \beta > 0$$
where $C$ is a normalizing constant that ensures the total probability integrates to 1 over the domain $x \geq 0$. To find $C$, we solve the integral,
$$\int_0^\infty C(1 + \beta x)^m e^{-m\beta x} dx = 1$$

Using the binomial expansion,
$$(1 + \beta x)^m = \sum_{k=0}^{m} \binom{m}{k} \beta^k x^k,$$
the integral becomes,
$$C \sum_{k=0}^{m} \binom{m}{k} \beta^k \int_0^\infty x^k e^{-m\beta x} dx = 1.$$

Recognizing that,
$$\int_0^\infty x^k e^{-m\beta x} dx = \frac{\Gamma(k+1)}{(m\beta)^{k+1}},$$
we obtain,
$$C \sum_{k=0}^{m} \binom{m}{k} \beta^k \frac{\Gamma(k+1)}{(m\beta)^{k+1}} = 1.$$

Simplifying,
$$C = \left( \frac{1}{\beta} \sum_{k=0}^{m} \binom{m}{k} \frac{\Gamma(k+1)}{m^{k+1}} \right)^{-1}.$$

Thus, the PDF corresponding to the detection function $g(x; m\, \beta)$ is,
$$f(x; m\, \beta) = \left( \frac{1}{\beta} \sum_{k=0}^{m} \binom{m}{k} \frac{\Gamma(k+1)}{m^{k+1}} \right)^{-1} (1 + \beta x)^m e^{-m\beta x}, x \geq 0, \beta > 0, m = 1,2,3,\ldots$$

The density at $x = 0$ (the parameter that we interest to estimate it) simplifies to,
$$f(0; m\, \beta) = \beta \left( \sum_{k=0}^{m} \binom{m}{k} \frac{\Gamma(k+1)}{m^{k+1}} \right)^{-1}$$

Table (1) below gives the simple formula of $f(0; m\, \beta)$ in terms of the parameter $\beta$ for different values of $m$. These expressions highlight how the value of the density at the origin changes with mmm, illustrating the influence of the parameter on the initial detectability.

**Table (1).** Closed-form expressions of $f(0; m\,\beta)$ for $m = 1, 2, 3, \ldots, 8$.

| $m$ | 1 | 2 | 3 | 4 | 5 | 6 | 7 | 8 |
|---|---|---|---|---|---|---|---|---|
| $f(0; m\,\beta)$ | $\dfrac{1}{2}\beta$ | $\dfrac{4}{5}\beta$ | $\dfrac{27}{26}\beta$ | $\dfrac{128}{103}\beta$ | $\dfrac{3125}{2194}\beta$ | $\dfrac{1944}{1223}\beta$ | $\dfrac{823543}{472730}\beta$ | $\dfrac{1048576}{556403}\beta$ |

## 4. Some Statistical Properties of $f(x; m\,\beta)$ and Estimation

Let $X$ be a ransom variable with PDF $f(x; m\,\beta)$. The r-th moment of $X$ is given by,

$$EX^r = \int_0^\infty C x^r (1+\beta x)^m e^{-m\beta x} dx.$$

Expanding using the binomial theorem,

$$= C \sum_{k=0}^{m} \binom{m}{k} \beta^k \int_0^\infty x^{k+r} e^{-m\beta x} dx.$$

This integral is the gamma integral, yielding,

$$= \frac{C}{\beta^{r+1}} \sum_{k=0}^{m} \binom{m}{k} \frac{\Gamma(k+r+1)}{m^{k+r+1}}.$$

Therefore, the general expression for the $r-th$ moment is,

$$EX^r = \frac{\sum_{k=0}^{m} \binom{m}{k} \frac{\Gamma(k+r+1)}{m^{k+r+1}}}{\beta^r \sum_{k=0}^{m} \binom{m}{k} \frac{\Gamma(k+1)}{m^{k+1}}}.$$

### 4.1 Mean, Variance and Moment Generation Function

From the expression for $EX^r$, we can derive the mean and variance directly. The mean of $X$ is,

$$EX = \frac{\sum_{k=0}^{m} \binom{m}{k} \frac{\Gamma(k+2)}{m^{k+2}}}{\beta \sum_{k=0}^{m} \binom{m}{k} \frac{\Gamma(k+1)}{m^{k+1}}},$$

and the variance is,

$$var(X) = \frac{\sum_{k=0}^{m} \binom{m}{k} \frac{\Gamma(k+3)}{m^{k+3}}}{\beta^2 \sum_{k=0}^{m} \binom{m}{k} \frac{\Gamma(k+1)}{m^{k+1}}} - \left( \frac{\sum_{k=0}^{m} \binom{m}{k} \frac{\Gamma(k+2)}{m^{k+2}}}{\beta \sum_{k=0}^{m} \binom{m}{k} \frac{\Gamma(k+1)}{m^{k+1}}} \right)^2.$$

The moment generating function (MGF) of $X$ can be derived as follows,

$$M_X(t) = E e^{tX} = C \int_0^\infty e^{tx} (1+\beta x)^m e^{-m\beta x} dx.$$

Expanding the binomial and integrating,

$$M_X(t) = C \sum_{k=0}^{m} \binom{m}{k} \beta^k \int_0^\infty x^k e^{tx} e^{-m\beta x} dx = C \sum_{k=0}^{m} \binom{m}{k} \beta^k \int_0^\infty x^k e^{-x(m\beta - t)} dx$$

$$= C \sum_{k=0}^{m} \binom{m}{k} \beta^k \frac{\Gamma(k+1)}{(m\beta-t)^{k+1}} = \frac{C}{m\beta-t} \sum_{k=0}^{m} \binom{m}{k} \Gamma(k+1) \left(\frac{\beta}{m\beta-t}\right)^k.$$

### 4.2 Estimation of $\beta$ and $f(0; m\beta)$ via Method of Moments

Assuming $m$ is known, we equate the theoretical mean $EX$ to the sample mean $\bar{X}$ and solve for $\beta$. This gives the method of moments estimator,

$$\hat{\beta} = \frac{1}{\bar{X}} \frac{\sum_{k=0}^{m} \binom{m}{k} \frac{\Gamma(k+2)}{m^{k+2}}}{\sum_{k=0}^{m} \binom{m}{k} \frac{\Gamma(k+1)}{m^{k+1}}}.$$

Using $\hat{\beta}$, we estimate the density at 0 as,

$$\hat{f}(0; m, \hat{\beta}) = \frac{1}{\bar{X}} \left(\sum_{k=0}^{m} \binom{m}{k} \frac{\Gamma(k+1)}{m^{k+1}}\right)^{-2} \left(\sum_{k=0}^{m} \binom{m}{k} \frac{\Gamma(k+2)}{m^{k+2}}\right).$$

Define the constant $L_m$ as,

$$L_m = \left(\sum_{k=0}^{m} \binom{m}{k} \frac{\Gamma(k+1)}{m^{k+1}}\right)^{-2} \left(\sum_{k=0}^{m} \binom{m}{k} \frac{\Gamma(k+2)}{m^{k+2}}\right),$$

then

$$\hat{f}(0; m\,\hat{\beta}) = \frac{L_m}{\bar{X}}.$$

Table (2) below presents the values of $L_m$ for different values of $m$. As $m$ increases, it is evident that $L_m$ steadily approaches a constant value, stabilizing around 0.6370.

**Table (2).** The values of $L_m$ for various of $m$.

| $m$ | 1 | 2 | 3 | 4 | 5 | 6 | 7 | 8 |
|---|---|---|---|---|---|---|---|---|
| $L_m$ | 0.75 | 0.72 | 0.7056 | 0.6978 | 0.6906 | 0.6860 | 0.6824 | 0.6795 |
| $m$ | 20 | 200 | 500 | 1000 | 5000 | 20000 | 50000 | 100000 |
| $L_m$ | 0.6638 | 0.6452 | 0.6420 | 0.6404 | 0.6383 | 0.6375 | 0.6372 | 0.6370 |

### 4.3 Estimation of β via Maximum Likelihood ($m$ is known)

We consider the estimation of the parameter $\beta$ in the detection model with known parameter $m \in N^+$. Let $X_1, X_2, \ldots, X_n$ be independent and identically distributed observations from the distribution $f(x; m, \beta)$. Define

$$C(m) = \left(\sum_{k=0}^{m} \binom{m}{k} \frac{\Gamma(k+1)}{m^{k+1}}\right)^{-1},$$

then the likelihood function for $\beta$, given the sample, is,

$$L(\beta) = \prod_{i=1}^{n} f(x_i; m, \beta) = \beta^n C(m)^n \prod_{i=1}^{n} (1+\beta x_i)^m e^{-m\beta x_i}.$$

The log-likelihood function is then,

$$logL(\beta) = nlog\beta + m\sum_{i=1}^{n} log(1 + \beta x_i) - m\beta \sum_{i=1}^{n} x_i + nlogC(m).$$

To obtain the maximum likelihood estimator (MLE) of $\beta$, we differentiate the log-likelihood function with respect to $\beta$,

$$\frac{dlogL(\beta)}{d\beta} = \frac{n}{\beta} + m\sum_{i=1}^{n} \frac{x_i}{1 + \beta x_i} - m\sum_{i=1}^{n} x_i.$$

Setting this equal to zero gives the likelihood equation

$$\frac{n}{m\beta} + \sum_{i=1}^{n} \frac{x_i}{1 + \beta x_i} = \sum_{i=1}^{n} x_i.$$

This equation does not have a closed-form solution for $\beta$, and therefore the MLE must be obtained numerically using root-finding algorithms such as the Newton-Raphson method or other optimization routines. Once the MLE $\hat{\beta}_{ML}$ is obtained, the corresponding estimate of the density at zero is $\hat{f}(0; m, \hat{\beta}_{ML})$. This provides a likelihood-based alternative to the method-of-moments estimator for $f(0)$.

## 5. Simulation Design

A simulation study was conducted to evaluate the performance of the proposed estimator $\hat{f}(0; m, \hat{\beta}) = L_m/\bar{X}$ under various distributional scenarios. Since the simulation results showed very similar behavior and conclusions for both the moment-based estimator $\hat{f}(0; m, \hat{\beta})$ and the maximum likelihood estimator $\hat{f}(0; m, \hat{\beta}_{ML})$, we present results only for the former estimator.

Three values of the tuning parameter $m$= 3, 8, 20 were examined. For comparison, two standard estimators frequently used in practice were included: the exponential estimator $\hat{f}_{EX}(0)$ and the half normal estimator $\hat{f}_{HN}(0)$. To simulate realistic and diverse shapes of detection functions, data were generated from four distinct families of distributions:

- Exponential Power (EP) Distribution ([8])
$$f(x) = \frac{1}{\Gamma(1 + 1/\delta)} e^{-x^\delta}, \quad x \geq 0, \quad \delta \geq 1$$
- Hazard-Rate (HR) Distribution ([42])
$$f(x) = \frac{1}{\Gamma(1 - 1/\delta)} \left(1 - e^{-x^{-\delta}}\right), \quad x \geq 0, \quad \delta > 1$$
- Beta (BE) Distribution ([43])
$$f(x) = (1 + \delta)(1 - x)^\delta, \quad 0 \leq x < 1, \quad \delta \geq 0$$
- General Polynomial (GP) Distribution ([12])
$$f(x) = \frac{10\Gamma(\delta)}{3\sqrt{\pi}\Gamma(\delta - 1/2)} (1 + (x/0.6)^2)^{-\delta}, \quad x \geq 0, \quad \delta > 1/2.$$

Each distribution was tested under three parameter settings, yielding 12 total configurations:

- EP distribution: $\delta = 1, \ 1.5, \ 2$
- HR distribution: $\delta = 2, \ 2.5, \ 3$
- BE distribution: $\delta = 1.5, \ 2, \ 2.5$
- GP distribution: $\delta = 0.9, \ 1.9, \ 3.5$

These combinations were selected to ensure a thorough evaluation of estimator behavior under a wide range of density shapes, from sharply peaked to nearly uniform. It is worth noting that the EP model with $\delta = 1$, along with all variations of the BE model, do not satisfy the shoulder condition, a common assumption in detection function modeling. These were deliberately included to test the robustness of the estimators under model misspecification. All other models conform to the shoulder condition.

For each of the 12 target models, random samples of perpendicular distances were generated for three sample sizes, $n = 50, 100$ and $200$. A total of 1,000 replications ($R = 1000$) were conducted for each scenario. Estimator performance was evaluated using the following metrics:

- Relative bias ($RB$)

$$RB = \frac{\hat{E}\left(\hat{f}(0)\right) - f(0)}{f(0)}.$$

- Relative root mean error ($RRMSE$)

$$RRMSE = \frac{\sqrt{\widehat{MSE}\left(\hat{f}(0)\right)}}{f(0)}.$$

- Efficiency (EFF), relative to the exponential estimator $\hat{f}_{EX}(0)$,

$$EFF = \frac{MSE(\hat{f}_{EX}(0))}{MSE(\hat{f}_U(0))},$$

where $U$ represents the estimator under evaluation.

## 6. Simulation Results

The simulation results reveal several key findings about the behavior and relative performance of the proposed estimator, which can be listed as follows:

1. Across almost all scenarios, the proposed estimator $\hat{f}(0; m, \hat{\beta})$ with moderate values of $m$ (particularly, $m = 8$) showed superior performance compared to the exponential and half-normal estimators in terms of lower RRMSE and higher relative efficiency.
2. For small $m$ (e.g., $m = 2$), the proposed estimator tended to have slightly higher bias and RRMSE, especially in light-tailed or sharply peaked distributions (e.g., EP with $\delta = 1$). As $m$ increased to 8 or 20, the estimator stabilized and consistently approached the true $f(0)$, achieving better bias-variance trade-offs.
3. Under the EP distribution with $\delta = 1$, none of the estimators performed optimally due to the violation of the shoulder condition. However, the proposed estimators still outperformed the half-normal estimator in both RB and RRMSE. For $\delta = 1.5$, the proposed estimator $\hat{f}(0; m, \hat{\beta})$ significantly outperformed both competitors, particularly for larger $n$, with EFF values often exceeding 4, indicating substantial gains in efficiency. It is important to note that when $\delta = 1$, the EP distribution coincides with the exponential distribution, where the exponential estimator $\hat{f}_{EX}(0)$ is naturally optimal. Similarly, for $\delta = 2$, the EP distribution corresponds to the half normal case, and thus $\hat{f}_{HN}(0)$ exhibits the best performance in this setting.
4. Under the HR distribution, with the exception of the case $\delta = 2$, the proposed estimators consistently achieved lower RB and RRMSE compared to both the exponential and half-normal estimators. For higher $\delta$ values, the efficiency gains were pronounced, particularly for $m = 8$ and $m = 20$, with EFF values frequently above 4. Although the HR distribution satisfies the shoulder condition when $\delta = 2$, its density decreases sharply as the perpendicular distance $x$ increases from zero. This rapid decline in ly as perpendicular distance increases away zero. This rapid decline in detection probability away from the origin can negatively impact the performance of estimators near $x = 0$.
5. Under the Beta (BE) distribution, which does not satisfy the shoulder condition, the proposed estimator exhibited notable robustness and adaptability, even in the presence of model misspecification. It consistently achieved a significantly lower relative root mean square error (RRMSE) compared to both the exponential estimator $\hat{f}_{EX}(0)$ and the half-normal estimator $\hat{f}_{HN}(0)$, particularly for $m = 2$ and $m = 8$. Moreover, it frequently outperformed these traditional estimators in terms of efficiency.
6. Under the GP distribution, the proposed estimator once again demonstrated strong performance, particularly for larger values of $\delta$ (e.g., $\delta = 3.5$), where it achieved both low RB and high efficiency. In scenarios with smaller $\delta$ values, such as $\delta = 0.9$, where the density is flatter and more dispersed, the estimator's performance was somewhat less optimal but remained competitive relative to the traditional alternatives.
7. The exponential estimator $\hat{f}_{EX}(0)$ consistently exhibited positive bias across all settings, particularly for non-exponential data. The half-normal estimator $\hat{f}_{HN}(0)$ was highly sensitive to model shape and often underperformed under misspecification, exhibiting large negative biases.
8. In general, as expected, increasing the sample size improved the performance of all estimators. However, the gains for the proposed estimator were more significant, especially in terms of reduced RRMSE and increased EFF, highlighting its favorable asymptotic properties.

In conclusion, across most distributional scenarios and sample sizes, the proposed estimator with $m = 8$ offered the best overall performance, striking a balance between bias reduction and variance control.

**Table (3).** The RB, RMSE and EFF for the estimators $\hat{f}_{EX}(0)$, $\hat{f}_{HN}(0)$, and $\hat{f}(0; m, \hat{\beta})$ with $m = 2, 8, 20$. The data are simulated from **Exponential Power (EP)** model.

| $\delta$ | $n$ | | $\hat{f}_{EX}(0)$ | $\hat{f}_{HN}(0)$ | $\hat{f}(0; m, \hat{\beta})$ | | |
|---|---|---|---|---|---|---|---|
| | | | | | $m = 2$ | $m = 8$ | $m = 20$ |
| 1 | 50 | RB | 0.025 | -0.414 | -0.262 | -0.303 | -0.319 |
| | | RRMSE | 0.148 | 0.424 | 0.282 | 0.319 | 0.334 |
| | | **EFF** | **1.000** | **0.348** | **0.524** | **0.463** | **0.442** |
| | 100 | RB | 0.013 | -0.425 | -0.270 | -0.311 | -0.327 |
| | | RRMSE | 0.107 | 0.430 | 0.281 | 0.320 | 0.335 |
| | | **EFF** | **1.000** | **0.249** | **0.381** | **0.335** | **0.320** |
| | 200 | RB | 0.006 | -0.430 | -0.276 | -0.316 | -0.332 |
| | | RRMSE | 0.074 | 0.432 | 0.281 | 0.320 | 0.336 |
| | | **EFF** | **1.000** | **0.171** | **0.263** | **0.231** | **0.220** |
| 1.5 | 50 | RB | 0.383 | -0.147 | -0.005 | -0.061 | -0.082 |
| | | RRMSE | 0.416 | 0.177 | 0.118 | 0.126 | 0.136 |
| | | **EFF** | **1.000** | **2.355** | **3.535** | **3.291** | **3.060** |
| | 100 | RB | 0.376 | -0.154 | -0.009 | -0.065 | -0.087 |
| | | RRMSE | 0.394 | 0.170 | 0.085 | 0.103 | 0.116 |
| | | **EFF** | **1.000** | **2.317** | **4.644** | **3.833** | **3.380** |
| | 200 | RB | 0.373 | -0.158 | -0.012 | -0.067 | -0.089 |
| | | RRMSE | 0.381 | 0.165 | 0.059 | 0.086 | 0.103 |
| | | **EFF** | **1.000** | **2.306** | **6.505** | **4.419** | **3.690** |
| 2 | 50 | RB | 0.583 | 0.013 | 0.140 | 0.076 | 0.051 |
| | | RRMSE | 0.607 | 0.104 | 0.185 | 0.137 | 0.123 |
| | | **EFF** | **1.000** | **5.861** | **3.278** | **4.420** | **4.940** |
| | 100 | RB | 0.585 | 0.011 | 0.141 | 0.077 | 0.052 |
| | | RRMSE | 0.597 | 0.072 | 0.165 | 0.111 | 0.094 |
| | | **EFF** | **1.000** | **8.300** | **3.624** | **5.378** | **6.400** |
| | 200 | RB | 0.575 | 0.0042 | 0.134 | 0.070 | 0.045 |
| | | RRMSE | 0.581 | 0.0501 | 0.147 | 0.091 | 0.072 |
| | | **EFF** | **1.000** | **11.600** | **3.951** | **6.407** | **8.000** |

**Table (4).** The RB, RMSE and EFF for the estimators $\hat{f}_{EX}(0)$, $\hat{f}_{HN}(0)$, and $\hat{f}(0; m, \hat{\beta})$ with $m = 2, 8, 20$. The data are simulated from **Hazard-Rate (HR)** model.

| $\delta$ | $n$ | | $\hat{f}_{EX}(0)$ | $\hat{f}_{HN}(0)$ | $\hat{f}(0; m, \hat{\beta})$ | | |
|---|---|---|---|---|---|---|---|
| | | | | | $m = 2$ | $m = 8$ | $m = 20$ |
| 2 | 50 | RB | 0.146 | -0.409 | -0.175 | -0.221 | -0.239 |
| | | RRMSE | 0.254 | 0.427 | 0.230 | 0.263 | 0.276 |
| | | **EFF** | **1.000** | **0.595** | **1.103** | **0.967** | **0.920** |
| | 100 | RB | 0.130 | -0.425 | -0.186 | -0.232 | -0.250 |
| | | RRMSE | 0.192 | 0.433 | 0.213 | 0.251 | 0.267 |
| | | **EFF** | **1.000** | **0.445** | **0.906** | **0.766** | **0.720** |
| | 200 | RB | 0.116 | -0.435 | -0.197 | -0.242 | -0.259 |
| | | RRMSE | 0.152 | 0.438 | 0.209 | 0.251 | 0.268 |
| | | **EFF** | **1.000** | **0.348** | **0.728** | **0.607** | **0.570** |
| 2.5 | 50 | RB | 0.405 | -0.217 | 0.012 | -0.045 | -0.067 |
| | | RRMSE | 0.463 | 0.265 | 0.161 | 0.158 | 0.163 |
| | | **EFF** | **1.000** | **1.747** | **2.869** | **2.922** | **2.840** |
| | 100 | RB | 0.378 | -0.240 | -0.008 | -0.063 | -0.085 |
| | | RRMSE | 0.404 | 0.258 | 0.102 | 0.115 | 0.126 |
| | | **EFF** | **1.000** | **1.564** | **3.976** | **3.520** | **3.200** |
| | 200 | RB | 0.378 | -0.245 | -0.008 | -0.064 | -0.085 |
| | | RRMSE | 0.393 | 0.256 | 0.078 | 0.097 | 0.111 |
| | | **EFF** | **1.000** | **1.538** | **5.037** | **4.048** | **3.530** |
| 3 | 50 | RB | 0.575 | -0.062 | 0.134 | 0.071 | 0.046 |
| | | RRMSE | 0.614 | 0.165 | 0.205 | 0.162 | 0.150 |
| | | **EFF** | **1.000** | **3.715** | **2.996** | **3.785** | **4.100** |
| | 100 | RB | 0.562 | -0.081 | 0.124 | 0.061 | 0.037 |
| | | RRMSE | 0.581 | 0.135 | 0.164 | 0.118 | 0.105 |
| | | **EFF** | **1.000** | **4.316** | **3.537** | **4.912** | **5.500** |
| | 200 | RB | 0.551 | -0.093 | 0.117 | 0.054 | 0.030 |
| | | RRMSE | 0.561 | 0.119 | 0.139 | 0.089 | 0.075 |
| | | **EFF** | **1.000** | **4.713** | **4.042** | **6.304** | **7.500** |

**Table (5).** The RB, RMSE and EFF for the estimators $\hat{f}_{EX}(0)$, $\hat{f}_{HN}(0)$, and $\hat{f}(0; m, \hat{\beta})$ with $m = 2, 8, 20$. The data are simulated from **Beta (BE)** model.

| $\delta$ | $n$ | | $\hat{f}_{EX}(0)$ | $\hat{f}_{HN}(0)$ | $\hat{f}(0; m, \hat{\beta})$ | | |
|---|---|---|---|---|---|---|---|
| | | | | | $m = 2$ | $m = 8$ | $m = 20$ |
| 1.5 | 50 | RB | 0.415 | -0.095 | 0.019 | -0.038 | -0.061 |
| | | RRMSE | 0.442 | 0.126 | 0.110 | 0.109 | 0.117 |
| | | **EFF** | **1.000** | **3.504** | **4.022** | **4.052** | **3.790** |
| | 100 | RB | 0.411 | -0.0973 | 0.016 | -0.041 | -0.063 |
| | | RRMSE | 0.425 | 0.114 | 0.078 | 0.083 | 0.094 |
| | | **EFF** | **1.000** | **3.737** | **5.477** | **5.148** | **4.510** |
| | 200 | RB | 0.401 | -0.103 | 0.009 | -0.048 | -0.070 |
| | | RRMSE | 0.408 | 0.111 | 0.054 | 0.069 | 0.085 |
| | | **EFF** | **1.000** | **3.681** | **7.620** | **5.893** | **4.780** |
| 2 | 50 | RB | 0.355 | -0.144 | -0.025 | -0.080 | -0.101 |
| | | RRMSE | 0.384 | 0.166 | 0.109 | 0.128 | 0.140 |
| | | **EFF** | **1.000** | **2.312** | **3.534** | **3.007** | **2.740** |
| | 100 | RB | 0.338 | -0.154 | -0.036 | -0.091 | -0.112 |
| | | RRMSE | 0.355 | 0.165 | 0.086 | 0.117 | 0.133 |
| | | **EFF** | **1.000** | **2.154** | **4.135** | **3.047** | **2.680** |
| | 200 | RB | 0.333 | -0.158 | -0.040 | -0.094 | -0.115 |
| | | RRMSE | 0.340 | 0.163 | 0.065 | 0.106 | 0.124 |
| | | **EFF** | **1.000** | **2.088** | **5.272** | **3.225** | **2.740** |
| 2.5 | 50 | RB | 0.299 | -0.187 | -0.065 | -0.117 | -0.138 |
| | | RRMSE | 0.332 | 0.204 | 0.122 | 0.153 | 0.168 |
| | | **EFF** | **1.000** | **1.629** | **2.719** | **2.175** | **1.980** |
| | 100 | RB | 0.289 | -0.194 | -0.072 | -0.124 | -0.145 |
| | | RRMSE | 0.307 | 0.203 | 0.105 | 0.144 | 0.161 |
| | | **EFF** | **1.000** | **1.514** | **2.934** | **2.142** | **1.910** |
| | 200 | RB | 0.283 | -0.199 | -0.076 | -0.128 | -0.148 |
| | | RRMSE | 0.292 | 0.203 | 0.092 | 0.137 | 0.156 |
| | | **EFF** | **1.000** | **1.438** | **3.159** | **2.124** | **1.870** |

**Table (6).** The RB, RMSE and EFF for the estimators $\hat{f}_{EX}(0)$, $\hat{f}_{HN}(0)$, and $\hat{f}(0; m, \hat{\beta})$ with $m = 2, 8, 20$. The data are simulated from **General Polynomial (GP)** model.

| $\delta$ | $n$ | | $\hat{f}_{EX}(0)$ | $\hat{f}_{HN}(0)$ | $\hat{f}(0; m, \hat{\beta})$ | | |
|---|---|---|---|---|---|---|---|
| | | | | | $m = 2$ | $m = 8$ | $m = 20$ |
| 0.9 | 50 | RB | 0.189 | -0.295 | -0.144 | -0.192 | -0.211 |
| | | RRMSE | 0.244 | 0.307 | 0.182 | 0.219 | 0.234 |
| | | **EFF** | **1.000** | **0.794** | **1.341** | **1.114** | **1.040** |
| | 100 | RB | 0.180 | -0.303 | -0.150 | -0.198 | -0.217 |
| | | RRMSE | 0.211 | 0.308 | 0.170 | 0.212 | 0.229 |
| | | **EFF** | **1.000** | **0.685** | **1.241** | **0.997** | **0.920** |
| | 200 | RB | 0.170 | -0.309 | -0.157 | -0.205 | -0.223 |
| | | RRMSE | 0.186 | 0.311 | 0.166 | 0.211 | 0.228 |
| | | **EFF** | **1.000** | **0.597** | **1.119** | **0.881** | **0.810** |
| 1.9 | 50 | RB | 0.281 | -0.263 | -0.077 | -0.129 | -0.149 |
| | | RRMSE | 0.336 | 0.292 | 0.153 | 0.179 | 0.193 |
| | | **EFF** | **1.000** | **1.151** | **2.197** | **1.872** | **1.740** |
| | 100 | RB | 0.272 | -0.276 | -0.084 | -0.136 | -0.155 |
| | | RRMSE | 0.301 | 0.290 | 0.125 | 0.161 | 0.177 |
| | | **EFF** | **1.000** | **1.039** | **2.408** | **1.867** | **1.700** |
| | 200 | RB | 0.270 | -0.284 | -0.086 | -0.137 | -0.157 |
| | | RRMSE | 0.283 | 0.289 | 0.107 | 0.150 | 0.168 |
| | | **EFF** | **1.000** | **0.979** | **2.659** | **1.893** | **1.690** |
| 3.5 | 50 | RB | 0.453 | -0.114 | 0.046 | -0.013 | -0.036 |
| | | RRMSE | 0.489 | 0.170 | 0.141 | 0.126 | 0.128 |
| | | **EFF** | **1.000** | **2.883** | **3.475** | **3.874** | **3.830** |
| | 100 | RB | 0.433 | -0.136 | 0.032 | -0.026 | -0.049 |
| | | RRMSE | 0.450 | 0.162 | 0.094 | 0.087 | 0.0947 |
| | | **EFF** | **1.000** | **2.781** | **4.808** | **5.160** | **4.750** |
| | 200 | RB | 0.425 | -0.142 | 0.026 | -0.032 | -0.054 |
| | | RRMSE | 0.434 | 0.155 | 0.070 | 0.069 | 0.081 |
| | | **EFF** | **1.000** | **2.795** | **6.235** | **6.307** | **5.400** |

## 7. Analysis of Real Data

The application of the proposed estimators to real data offers valuable insights into their practical utility. It enhances understanding of their empirical performance, supports further methodological refinement, and helps determine suitable contexts for their effective use. In this section, we reanalyze the Laake stakes dataset, originally presented in Burnham et al. [1], by applying the proposed estimator,

$$\hat{f}(0; m, \hat{\beta}) = \frac{L_m}{\bar{X}}$$

for various values of $m$: 1, 2, 3, 8, 20, and 100,000. For comparison, we also consider the exponential estimator $\hat{f}_{EX}(0)$ and the half-normal estimator $\hat{f}_{HN}(0)$.

The data were collected using the line transect method to estimate the density of stakes. Perpendicular distances (in meters) from the transect line to 68 detected stakes were recorded out of a total of 150 stakes randomly distributed over an area of length $L = 1000$ meters. The true density of stakes is known to be $D = 0.00375$ stakes/m² (or 37.5 stakes/hectare), and this gives the true value of $f(0) \approx 0.110294$, calculated from the formula,

$$D = \frac{nf(0)}{2L}.$$

Table (7) presents the point estimates, bootstrap means, biases, standard deviations (SD), and mean squared errors (MSE) of each estimator. The sample mean of the 68 perpendicular distances is 6.10824 meters. The bias, SD and MSE for each estimator were computed using the bootstrap method with 1000 replications ($R = 1000$).

The exponential estimator $\hat{f}_{EX}(0)$ = 0.16371 substantially overestimates $f(0)$, with a relatively high MSE of 0.00035. The half-normal estimator $\hat{f}_{HN}(0) = 0.09752$ underestimates $f(0)$, but performs better than the exponential estimator, with a lower MSE of 0.00017.

Among the proposed estimators $\hat{f}(0; m, \hat{\beta})$, the estimate improves as $m$ increases. For smaller $m$ values (1, 2, 3), the estimators tend to slightly overestimate $f(0)$, while for larger $m$, especially $m = 8$ and $m = 20$, the estimates closely approach the true value of 0.110294. At $m = 8$, the estimator yields $\hat{f}(0; m, \hat{\beta}) = 0.11124$, which is the closest to the true value and also has the lowest MSE among all evaluated estimators, at 0.00016.

When $m = 100,000$m=100, the estimator slightly underestimates $f(0)$, suggesting a limit to how beneficial increasing $m$ can be in practice. This estimator yields the lowest MSE (0.00014), but the estimate (0.10429) deviates slightly more from the true value than the $m = 8$ and $m = 20$ cases.

Corresponding estimates for density $D$ (Last column of Table 7) align with the behavior of the $f(0)$ estimates. Note that the proposed estimator with $m = 8$ gives $\hat{D} = 0.00378$, which is extremely close to the true value.

**Table (7).** Point estimates of $f(0)$ along with the corresponding bias, standard deviation (SD) and mean square error (MSE) for each estimator. The true value of $f(0)$ is 0.110294 and the corresponding true density $D$ is 0.00375.

|  | Estimate | Bootstrap Mean | Bias | SD | MSE | $\widehat{D} = n\hat{f}/2000$ |
|---|---|---|---|---|---|---|
| $\hat{f}_{EX}(0)$ | 0.16371 | 0.16596 | 0.00224 | 0.01859 | 0.00035 | 0.00557 |
| $\hat{f}_{HN}(0)$ | 0.09752 | 0.09982 | 0.00223 | 0.01288 | 0.00017 | 0.00332 |
| $\hat{f}(0; m, \hat{\beta})$ $m = 1$ | 0.12279 | 0.12447 | 0.00168 | 0.01394 | 0.00020 | 0.00417 |
| $\hat{f}(0; m, \hat{\beta})$ $m = 2$ | 0.11787 | 0.11949 | 0.00162 | 0.01338 | 0.00018 | 0.00401 |
| $\hat{f}(0; m, \hat{\beta})$ $m = 3$ | 0.11552 | 0.11710 | 0.00158 | 0.01312 | 0.00018 | 0.00393 |
| $\hat{f}(0; m, \hat{\beta})$ $m = 8$ | 0.11124 | 0.11277 | 0.00152 | 0.01263 | 0.00016 | 0.00378 |
| $\hat{f}(0; m, \hat{\beta})$ $m = 20$ | 0.10867 | 0.11016 | 0.00149 | 0.01234 | 0.00015 | 0.00369 |
| $\hat{f}(0; m, \hat{\beta})$ $m = 100000$ | 0.10429 | 0.10571 | 0.00143 | 0.01184 | 0.00014 | 0.00355 |

## 8. Discussion and Conclusion

The results demonstrate that the proposed model $f(x; m, \beta)$, which leads to the estimator $\hat{f}(0; m, \hat{\beta})$ for $f(0)$ (a central quantity in line transect sampling) is a reliable and flexible model for estimating detection probabilities. It provides a clear advantage over classical estimators, particularly when moderate values of the tuning parameter $m$ are used. In practical applications, selecting m=8m = 8m=8 offers a robust and effective trade-off balance between bias and variance, making the proposed approach a promising and robust alternative for density estimation in line transect sampling used broadly in ecological and wildlife population studies.